\documentstyle[epsfig,sprocl]{article}

\def\be{\begin{equation}}
\def\ee{\end{equation}}
\def\bea{\begin{eqnarray}}
\def\eea{\end{eqnarray}}

\begin{document}

\title{Are electroweak corrections at 1 TeV \\
under control at the 1 \% level?}

\author{P. CIAFALONI }

\address{Dipartimento di Fisica \& INFN - sezione di Lecce\\
Via Arnesano, I-73100 Lecce}


\maketitle\abstracts{
Future lepton colliders will provide a powerful tool for making
precision experiments at energies that will range typically between 
 500 GeV and 2 TeV. At such high energies,
one loop electroweak corrections 
are bigger than one could na\"\i vely expect 
{\it a priori}. Thus, the calculation
of higher order  electroweak effects (and  possibly their resummation) 
might be needed. 
}
  
\section{Introduction}

High precision experiments at LEP have been able to prove the quantum
structure of the electroweak theory at the {\it per mille} 
level \cite{LEP}. These experiments  have tested the Standard Model (SM)
electroweak sector  at  energies close to the Z mass $M_Z\approx 91$ GeV. 
The typical magnitude of SM electroweak
corrections is dictated at LEP by the perturbative series expansion parameter
$\frac{\alpha(M_Z)}{4\sin^2\theta_w\pi}\approx 2.7\times 10^{-3}$, where 
$\alpha(M_Z)$ is the QED effective coupling constant at the energy $M_Z$ 
and  $\sin\theta_w$ is the Weinberg angle. Since the experimental accuracy is
of comparable magnitude, the well known one-loop electroweak corrections 
\cite{Zphys}  are sufficient in general to allow for a comparison between
theory and experiment. As an exception, some leading two loop electroweak
corrections \cite{noiedegrassi} 
growing with the top mass $m_t$ turn out to be also relevant at LEP.

While experiments at LEP have tested the electroweak theory at its
characteristic energy of about 100 GeV, future experiments will   
go much beyond this mass scale. This holds  in particular
for the generation of linear colliders \cite{NLC}. These colliders will
feature high luminosities, allowing for precision experiments 
at energies ranging from 500 GeV to 2 TeV.
The calculation of electroweak corrections at such high energies 
with a precision comparable to the experimental accuracy is then an 
important issue. Recent results \cite{double} seem to indicate that 
yet uncalculated higher order electroweak effects and/or their possible
resummation are indeed important for future linear colliders. 

\section{IR divergences and double logs}

What is the order of magnitude of electroweak corrections 
that one expects at a typical energy of, say, 1 TeV?  
Let us assume that we determine
the SM parameters with high precision through a series of 
LEP experiments at the Z mass. Then 
we expect that perturbative corrections for an observable measured at a
different c.m. energy $\sqrt{s}$ are enhanced by large logarithms 
of ultraviolet (UV) origin of the form 
$\frac{\alpha(M_Z)}{4\sin^2\theta_w\pi}\log\frac{s}{M_Z^2}
\approx 1.3 \times 10^{-2}$ for $\sqrt{s}=1$ TeV. 
Since the one loop effects are of the order of
1 \%, we expect higher order effects to be of the order of 0.1 \%. Moreover,
the large logarithms can be resummed at all orders through renormalization
group equations (RGEs). Then, if we fix the expected experimental accuracy to
be at the 1 \% level at NLCs (which is probably a conservative assumption
since the accuracy is expected to be better than this \cite{NLC}), there is no
need to worry at all: electroweak corrections are under control, i.e. they are
theoretically known through one-loop results with an accuracy which is better
than the experimental one. 

However, this is not the end of the story. 
As has already been noticed\cite{Kuroda}, 
electroweak corrections also contain terms growing 
with the energy $\sqrt{s}$
like the {\it square} of a log, i.e. proportional to 
$\log^2\frac{s}{M_{Z,W}^2}$. This can be understood as follows:
when the energy is much bigger than the mass of all the particles running in
the loops, which means $\sqrt{s}>>M_W,M_Z$ if we don't consider processes in
which the top quark plays a role, the W and Z mass act as an effective cutoff
for infrared (IR) divergences. Infrared divergences arise in
perturbative calculations from regions of integration over the loop 
momentum $k$ where $k$ is small compared to the typical scales of 
the process.  
This is a well known fact in QED for instance \cite{Landau} 
where the problem of an
unphysical divergence is solved by giving the photon a fictitious mass which
acts a a cutoff for the IR divergent integral.  When real (bremsstrahlung) and
virtual contributions are summed, the dependence on this mass cancels and the
final result is finite \cite{Landau}.  The (double) logarithms coming from
these contributions are large and, growing with the scale, can spoil
perturbation theory and need to be resumed. They are usually called {\rm
Sudakov} double logarithms \cite{Sudakov}.  In the case of electroweak
corrections, similar logarithms arise when the typical scale of the process
considered is much larger than the mass of the particles running in the loops,
typically the $W(Z)$ mass \cite{Kuroda,sirling,verza}.  The expansion
parameter results then
$\frac{\alpha}{4\sin^2\theta_w\pi}\log^2\frac{s}{M_W^2}$, which is already 
7 \% for for energies $\sqrt{s}$ of the order of 1 TeV. 
In the case of corrections coming from
loops with $W(Z)$s, there is no equivalent of ``bremsstrahlung'' like in QED
or QCD: the $W(Z)$, unlike the photon, has a definite nonzero mass and is
experimentally detected like a separate particle. In this way the full
dependence on the $W(Z)$ mass is retained in the corrections.  
Let us consider IR divergences coming from vertex corrections for instance
(also box diagrams are present in two fermion production, but I do not discuss
them here).
For simplicity, I consider a  ``SM-like case'' 
in which a ``W boson'' having mass $M$ and
coupling with fermions like the photon is exchanged. 
In the limit of massless fermions considered here, there is no coupling to the
Higgs sector. Moreover, by power counting arguments, it is easy to see that the
vertex correction where the trilinear gauge boson coupling appears is not IR
divergent. The only potentially IR divergent diagram is then the one of
fig. 1, where a gauge boson is exchanged in the t-channel. 
It is convenient to choose the  momentum of 
integration $k$ to be the one
 of the exchanged particle, the boson in this case. 
Then, by simple power
counting arguments it is easy to see that the IR divergence can only be
produced by regions of integration where $k\approx 0$.
The only potentially IR divergent
integral is then the scalar integral, 
usually called $C_0$ in the literature \cite{veltman}.
 Any other integral with $k_\mu,k_{\mu}k_{\nu}$ in the numerator
cannot, again
by power counting, be IR divergent.
To understand the origin of the divergences, let us consider the diagram
of fig.1
with all the masses  set to zero. 
For $k\approx 0$ the leading term of the vertex amplitude is given by:
\be\label{QED}
{\cal V}\approx 
-\frac{\alpha}{4 \pi}{\cal V}_0  \int\frac{d^4k}{i\pi^2}
\frac{(p_1p_2)}{k^2(kp_1)(kp_2)}
\approx
-\frac{\alpha}{2 \pi}{\cal V}_0 \int_0^1\frac{dx}{x}\int_0^{1-x}\frac{dy}{y}
\ee
where ${\cal V}_0$ is the tree level vertex. 
We can see here the two logarithmic
divergences that arise from the integration over the $x,y$ Feynmann
parameters. As is well known \cite{Landau}, one of them is of collinear
origin and the other one is a proper IR divergence. When we take some of the
external squared momenta and/or masses different from zero, they serve as
cutoffs for the divergences. The bottom line is that the Feynman diagram of
fig. 1 produces a term proportional to $\alpha\log^2\frac{s}{M^2}$, where M is
the exchanged boson mass.
It is easy to see that the dependence on the IR logs 
simply factorizes for the cross section 
($\sigma_B$ is the tree level cross section):
\[
\sigma\propto  \frac{1}{s}\int_0^s\frac{dt}{s}
|{\cal M}_0|^2 [1-2\frac{\alpha}{4 \pi} \log^2\frac{s}{M^2}]
=\sigma_B[1-2\frac{ \alpha}{4 \pi} \log^2\frac{s}{M^2}]
\]

\begin{figure}[htb]\setlength{\unitlength}{1cm}
\begin{center}
\begin{picture}(10,3)
\put(3,0){\epsfig{file=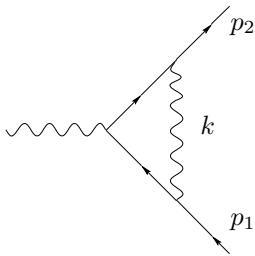,width=3cm}}
\put(5.6,1.6){$k$}
\put(6,3){$p_2$}
\put(6,.4){$p_1$}
\end{picture}
\end{center}
\caption{Vertex diagram  generating a  
$\log^2 \frac{s}{M^2}$. $p_1$ and $p_2$ are ingoing. }
\end{figure}

\section{Asymptotic behavior of two fermion processes}

In \cite{double} and \cite{singledouble}, 
the production of two massless fermions 
in an high energy lepton collider has been considered.
In \cite{double} the coefficients of the leading terms, growing with the energy
like the square of a log as we have seen, have been calculated. In
\cite{singledouble} also the coefficients of the subleading terms, growing like a
single log and of collinear origin, have been calculated for various
observables.  In the following, I call ``Sudakov-type'' logs the 
single and double logs of IR and
collinear origin, to distinguish them from the logs of UV origin.
 As an example, let me consider the total crossection for the
process $e^+e^-\to\mu^+\mu^-$. From \cite{singledouble}, we get:
\be\label{mu}
\sigma_\mu\stackrel{\sqrt{s}>>M}{\approx}
\sigma_B\;\; \{1+\frac{\alpha}{4\pi\sin^2\theta_w}
[0.6 \; L_{UV}+9.4 \; L_{IR}-1.4 \; L_{IR}^2]\}
\ee
Here, $ L_{UV}$ and $ L_{IR}$ are numerically the same:
 $ L_{UV}= L_{IR}=\log\frac{s}{M^2}$ and M is the weak scale
$M\approx M_W\approx M_Z\approx 91$ GeV. $\sigma_B$ is the Born cross section,
precisely defined in \cite{singledouble}. This formula is expected to describe
the full one loop calculation better and better as the energy grows, since the
subleading terms that have not been extracted become less and less important
with respect to the leading logarithmic ones. Indeed, formula \ref{mu} well
approximates the exact result coming from numerical programs \cite{topazo}
for energies around 1 TeV(see \cite{singledouble}). 
However, for energies well above 1
TeV where the agreement is supposed to be even better, no numerical
computation is available at the moment.
The graph corresponding to eq. \ref{mu} is drawn in fig. 2, where the relative
deviation for the total cross section
$\frac{\Delta\sigma}{\sigma}\equiv \frac{\sigma-\sigma_B}{\sigma_B}$
 is drawn as a function of the c.m. energy, and the various
contributions are also separately plotted. 

One evident feature of eq. \ref{mu} 
is that, while the coefficients of the single UV log and the
 double Sudakov log are of order 1, as one could expect {\it a priori}, 
the coefficent of the single Sudakov-type single log is of order  
10.
This has two immediate consequences:
\begin{itemize}
\item
The contribution of the single log of UV origin is almost negligible with
respect to the Sudakov logs ones. Thus, a na\"\i ve expectation of an
asymptotic behavior dictated by the UV structure of the theory turns out to be
wrong.
\item
Since the sign of the double and single Sudakov logs are opposite, there are
big cancellations and the correction to $\sigma_\mu$ crosses a zero at an
energy of about 2 TeV
\end{itemize}
These features are most easily seen by looking at fig. 2, where the net result
is seen to result from cancellations of big contributions  
of Sudakov single and double logs of opposite signs, while the RGE driven logs
are almost negligible.
\begin{figure}[htb]\setlength{\unitlength}{1cm}
\begin{center}
\begin{picture}(10,7)
\put(.3,7){\epsfig{file=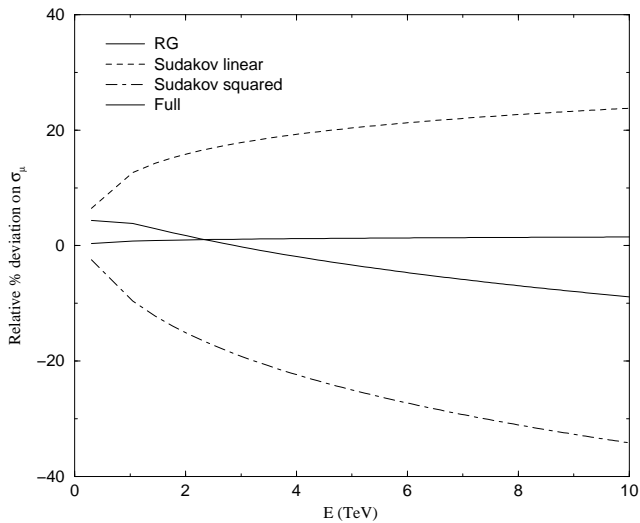,width=7cm,angle=270}}
\end{picture}
\end{center}
\caption{Relative deviation $\frac{\Delta\sigma}{\sigma}$ 
for the total cross section of $e^+e^-\to\mu^+\mu^-$ 
 }
\end{figure}

Where does all this leave us with the question posed with the title? In the
end, one could think that when doing perturbative calculations, big
cancellations can  {\it always} be 
present (between different graphs contributing
to the same amplitude for
instance). Then, since here the net effect is only a few percent in the
considered energy range as one can see from fig. 2, there is no need to worry
about higher order effects. However, the situation here is different in my
opininon. Here we have terms that are separately gauge invariant and have
different energy behavior. The fact that their contribution almost exactly
cancels at an energy of about 2 TeV which is close to the energy of interest,
is to be taken as accidental. 
Let us take another point of view: the relative effect of double logs is, from 
eq. 2, $1.4\frac{\alpha}{4\pi\sin^2\theta_w} \log^2\frac{s}{M^2}\approx
0.1$ at 1 TeV. A two loop calculation will produce a term growing like the
fourth power of a log, of the order $(0.1)^2=1$ \%. Until an higher order
calculation will be done, one cannot say that electroweak corrections at 1 TeV
are under control at the 1 \% level.

\end{document}